# Title: Pico-Newton mechanical forces promote neurite growth

**Authors:** V. Raffa[1,2*], F. Falcone[1,2], M. P. Calatayud[3], Gerardo F. Goya[3], Alfred Cuschieri[1]

**Affiliations:**

[1]The Institute for Medical Science and Technology, University of Dundee, Dundee DD1 4HN, UK

[2]Department of Biology, Università di Pisa, S.S. 12 Abetone e Brennero 4, Pisa 56127, IT

[3]Instituto de Nanociencia de Aragón, Universidad de Zaragoza, Mariano Esquillor, Zaragoza, Spain

*Correspondence to: vittoria.raffa@unipi.it

**Abstract**: Investigations over half a century have indicated that mechanical forces induce neurite growth - with neurites elongating at a rate of 0.1-0.3μmh$^{-1}$ per pico-Newton (pN) of applied force - when mechanical tension exceeds a threshold, with this being identified as 400-1000 pN for neurites of PC12 cells. Here we demonstrate that there is no threshold for neurite elongation of PC12 cells in response to applied mechanical forces. Instead this proceeds at the same previously identified rate, on application of tensions with intensity below 1pN. This supports the idea of mechanical tension as an endogenous signal used by neurons for promoting neurite elongation.

**Summary** Mechanical tension could be an endogenous signal used by neurons for promoting neurite growth. Raffa et al show that this stretched-growth can also occur at mechanical tensions sensibly lower than the force generated *in vivo* by axons and growth cones, supporting this idea.

## Introduction

With body growth in humans and large animals, the distance between the neuron soma and its cellular target increases, imposing stretch on neurites. Paul Weiss in 1941 hypothesized that the tensile force originating from this growth-induced stretching could be a signal that causes the neural processes to lengthen. Currently, it is widely accepted that neurites elongate, when mechanical tension exceeds a threshold, the process being referred to as "stretch-growth" (Franze and Guck, 2010; O'Toole et al., 2008a; Suter and Miller, 2011). The elongation rate was found to be very similar for both the central and peripheral nervous system (0.1-0.3μmh$^{-1}$pN$^{-1}$) (Chada et al., 1997; Fass and Odde, 2003; Zheng et al., 1991) but various thresholds have been identified. A force threshold for elongation of about 1nN was reported for neurites of PC12 cells (Lamoureux et al., 1997) and for chick sensory neurons (Zheng et al., 1991) elongated by the pulling force of glass microneedles, and 15-100pN in neurites of chick forebrain neurons (Fass and Odde, 2003) elongated by the magnetic force of magnetic microbeads. Traditionally, the addition of new cytoskeletal mass to the neurite was thought to occur at the leading edge, the growth cone. However, in stretch-growth, mass addition occurs at any site of increased tension, e.g. at the tip when the growth cone is pulled or along the whole neurite length when the entire neurite is stretched (Lamoureux et al., 2010; Miller and Sheetz, 2006; O'Toole et al., 2008b). Indeed, neuron could regulate neurite elongation at sites other than the growth cone (Ruthel and Hollenbeck, 2000).The 'stretch-growth model' was formulated, using these reported observations (O'Toole et al., 2008a). This postulated that mechanical tension may act, akin to a second messenger, as a regulator of neurite initiation and elongation; neurite elongation is driven by tension, independent of its origin, i.e. from the traction exerted by the growth cone, the mass body growth or an external applied force. Tip growth may be regarded as a special case of stretch-growth where the growth cone is responsible for creating the tension required for neurite elongation and mass addition occurs at the tip where this tension is localized (Lamoureux et al., 1989).

However, there is uncertainty that stretch-growth can realistically provide the basis for a unified model of neurite growth. The existence of a threshold makes unrealistic stretch-growth *in vivo*. For example, it has been reported that neurites of PC12 cells exhibit a transient elongation interpreted as viscoelastic deformation, when the applied tension is less than 0.4-1nN (Dennerll et al., 1988; Dennerll et al., 1989); in contrast, long-term extension, interpreted as growth is observed when the applied tension is above 0.4-1nN. However, some studies reported tensions of the order of 300–400 pN along PC12 neurites cultured *in vitro* (Dennerll et al., 1988), suggesting that the mechanical tension created at the growth cone is insufficient to trigger stretch-growth. Similar considerations apply also to central and peripheral nervous system neurons (Athamneh and Suter, 2015).

The present study assumed that the experimental approaches used in the past to identify the threshold for stretch-growth were methodologically suspected. In fact, the low detection limit (100pN for glass microneedles and 15pN for magnetic microbeads) together with the short observation periods (1 hour or less) used in previous studies were inadequate for studies on the effect of pN forces. On this premise, the present study aimed to investigate the effect of extremely low forces with a modern experimental set up, based on the use of superparamagnetic iron oxide nanoparticles (MNPs).

**Results and discussion**

MNPs can be safely administered to neurons. We have extensively tested these particles in neuronal cell lines, primary neurons and organotypic neuronal cultures (Calatayud et al., 2013; Calatayud et al., 2014; Pinkernelle et al., 2015; Riggio et al., 2014; Riggio et al., 2013). Collectively, our data on cell viability exclude any MNP-induced cell toxicity (supplementary information, S1). We have also extensively characterized cell-particle interactions by electron microscopy. MNP labelling consists in the addition of the particles to the medium. Particles usually stick to the cell surface as initial event (Calatayud et al., 2013; Calatayud et al., 2014; Riggio et al., 2014). Subsequently, they are avidly internalized by cells and the agglomerates occupy the intracellular space. Microanalysis performed on cross-sectioned cells confirmed the particle localization within cell cytoplasm. MNPs were found to be abundant in cell neurites (Riggio et al., 2014). Fig. 1 shows the localization of MNPs in cell neuritis. MNPs appear as electron-dense spots (white arrows) and the iron content is confirmed by microanalysis (Fig. 1A2). We demonstrated that the forces developed by MNPs entrapped in neurites, under the effect of magnetic fields (M), can be used to manipulate the neurites of differentiated PC12 cells (Riggio et al., 2014). In fact, MNPs were used to develop a tangential force against the neurites, which turned in response to this force, by preferentially aligning the direction of growth to the direction of the magnetic force (Riggio et al., 2014).

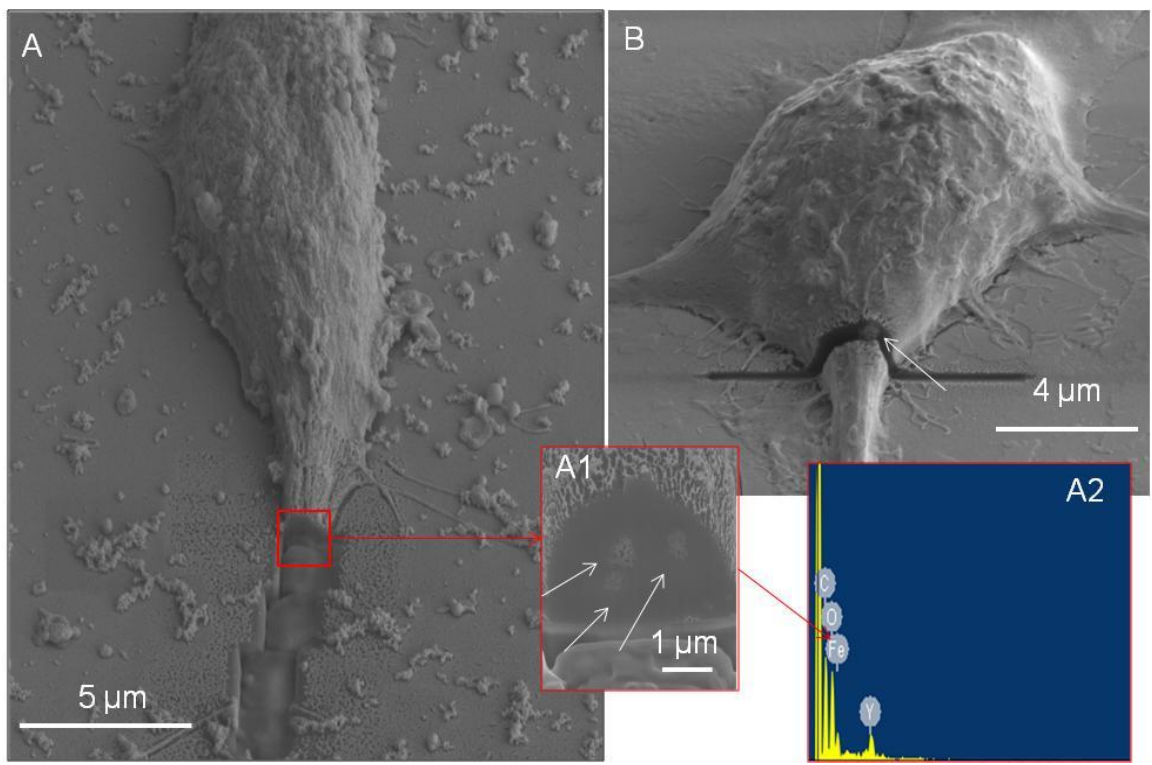

**Fig. 1**. SEM imaging and FIB milling of differentiated cells pre-incubated with 10µg/ml of MNPs. A-B) Cross-section of neurites showing electron-dense nanoparticles (pointed by white arrow).

Panel A2) shows Energy Dispersive X-ray (EDX) analysis of A1) inset, which confirms that the electron-dense nanoparticles cointain iron.

In the present study we used the force generated by MNPs in response to static magnetic fields for stretching the neurites. Differentiated MNP-labelled PC12 cells were exposed to a constant magnetic field that produces a magnetic force vector constant in amplitude, direction and orientation (Riggio et al., 2014). The mathematical model used (supplementary information, S2) predicts that, with the experimental conditions used, the MNPs exert a mechanical force on the neurite below 1nN (in the range of 0.4-0.9pN). The stretching time (72-120h) was chosen to produce neurite elongation at easily observable length of microns or tens of microns, in accordance with the elongation rate of 0.1-0.3µmh$^{-1}$ per pN of applied force, reported for PC12 cells (Lamoureux et al., 1997). The elongation analysis was carried on stretched cells (i.e. cells labelled with the particles and exposed to the magnetic field, hereafter labelled as M$^+$MNP$^+$) and on control groups, i.e. non-stretched cells treated with the same magnetic field (labelled as M$^+$MNP$^-$) or with the particles (labelled as M$^-$MNP$^+$) or untreated (labelled as M$^-$MNP$^-$). We tested 2 doses of MNPs (3.45pg or 4.85pg of MNPs per cell) and two stretching times (72h and 120h). The procedure of MNP labelling and the assessment of MNP amount per cell is described in supplementary information, S3. Each experiment was repeated 3 times with the experiments being blinded and performed after random allocation by 2 different operators. Table 1 provides each experiment (n=200) and the corresponding statistical analysis. Fig. 2.A1-3 plots overall data of 3 biological replicates (n=600) for each stretching condition.

**Table 1.** Neurite length (µm) for each experiment.

| Experiment description | Repl | M$^-$MNP$^-$ | M$^+$MNP$^-$ | M$^-$MNP$^+$ | M$^+$MNP$^+$ | P value |
|---|---|---|---|---|---|---|
| 3.4pg MNP, 72h stretch | R1 | 44.34±2.10 | 44.02±1.87 | 46.42±1.69 | 56.45±2.09***,###, §§§ | 6·10$^{-10}$ |
| | R2 | 41.22±1.95 | 39.17±1.43 | 40.47±1.60 | 52.13±2.37***, ###,§§§ | 2.6·10$^{-6}$ |
| | R3 | 45.80±1.86 | 44.41±2.00 | 44.75±2.12 | 56.16±2.78**, ###, §§§ | 2.7·10$^{-5}$ |
| 3.4pg MNP, 120h stretch | R1 | 55.82±2.36 | 59.83±3.39 | 63.78±3.56 | 84.69±4.37***, §§§, ### | 1.8·10$^{-10}$ |
| | R2 | 58.46±2.37 | 62.91±2.90 | 65.59±2.73 | 90.47±4.67***, ###, §§ | 2.8·10$^{-7}$ |
| | R3 | 62.78±2.82 | 58.04±2.73 | 63.03±2.93 | 86.58±4.16***, ###, §§§ | 1.2·10$^{-8}$ |
| 4.8pg MNP, 72h stretch | R1 | 46.76±2.12 | 49.38±2.08 | 48.93±2.12 | 64.03±3.07***,##, §§§ | 1.9·10$^{-5}$ |
| | R2 | 50.00± 1.82 | 47.67±2.45 | 50.72±2.39 | 66.66±3.09**, ###, §§§ | 5.5·10$^{-8}$ |
| | R3 | 45.21± 2.04 | 49.54±2.59 | 49.56±2.60 | 65.65±3.53***,##, §§ | 5.7·10$^{-5}$ |
| NGF 2h | R1 | 30.38±0.95 | 30.26±0.96 | 30.45±0.97 | 40.86±1.60***, ###, §§§ | 5.7·10$^{-11}$ |
| | R2 | 31.11±1.05 | 32.13±0.91 | 32.61±1.18 | 40.29±1.2***, ##, §§§ | 5.8·10$^{-7}$ |
| | R3 | 28.70±0.75 | 29.60±0.94 | 33.02±1.47 | 40.29±1.67***, ###, §§§ | 3.5·10$^{-14}$ |

N=200. Kruskal Wallis test followed by HDS correction. "*" is the significance vs. the control group (M$^-$MNP$^-$), "#" is the significance vs. the group treated with the magnet (M$^+$MNP$^-$) and "§"is the significance vs. the group treated with particles (M$^-$MNP$^+$)

In each experiment (Table 1) and in each stretching condition (Fig. 2.A1-3) tested, we found that the pN-stretching triggers a statistically highly significant increase ($p < 0.01$) in the length of neurites, when compared to any other group; whereas the control groups (M$^-$MNP$^-$, M$^-$MNP$^+$, M$^+$MNP$^-$) did not differ from each other ($p>0.05$). Collectively, these observations confirm excellent experimental reproducibility. They exclude any non-specific effect triggered by the particles or magnetic fields alone, and indicate that the stretch is responsible for the length increase. In order to demonstrate that the observed length increase is not a viscoelastic deformation but genuine growth, we calculated the average thickness of neurites. The analysis was performed in the experimental condition that yielded the highest differential elongation (3.4pg MNP, 120h). Although the stretched neurites were 44.44 ± 3.18% longer than control groups, there was no difference in neurite thickness among all the groups ($p=0.43$), indicating that the observed elongation was the result of actual growth due to mass addition (Fig. 2.B). Interestingly, we found the elongation rate to be constant ($0.252\pm0.005\mu mh^{-1}pN^{-1}$) (Fig. 2.C) with no difference between the three stretching conditions ($p=0.06$, n=3, 1-way ANOVA). It was very similar to the elongation rate calculated in previous reported studies (Dennerll et al., 1988; Dennerll et al., 1989), in which the applied force was 5 orders of magnitude higher than that used in the present study. Additionally, the present study provides an evidence that stretch-growth depends on force direction but not orientation, even if the elongation is oriented (from soma to tip), in line with the recent observation that mechanical forces can also drive retrograde axon extension (Breau et al., 2017). Specifically, stretch-growth depends on the longitudinal component of the mechanical force (with respect to the neurite direction) but both orientations (i.e. from tip to soma or from soma to tip) of the force produce stretch-growth (supplementary information, Fig. S2).

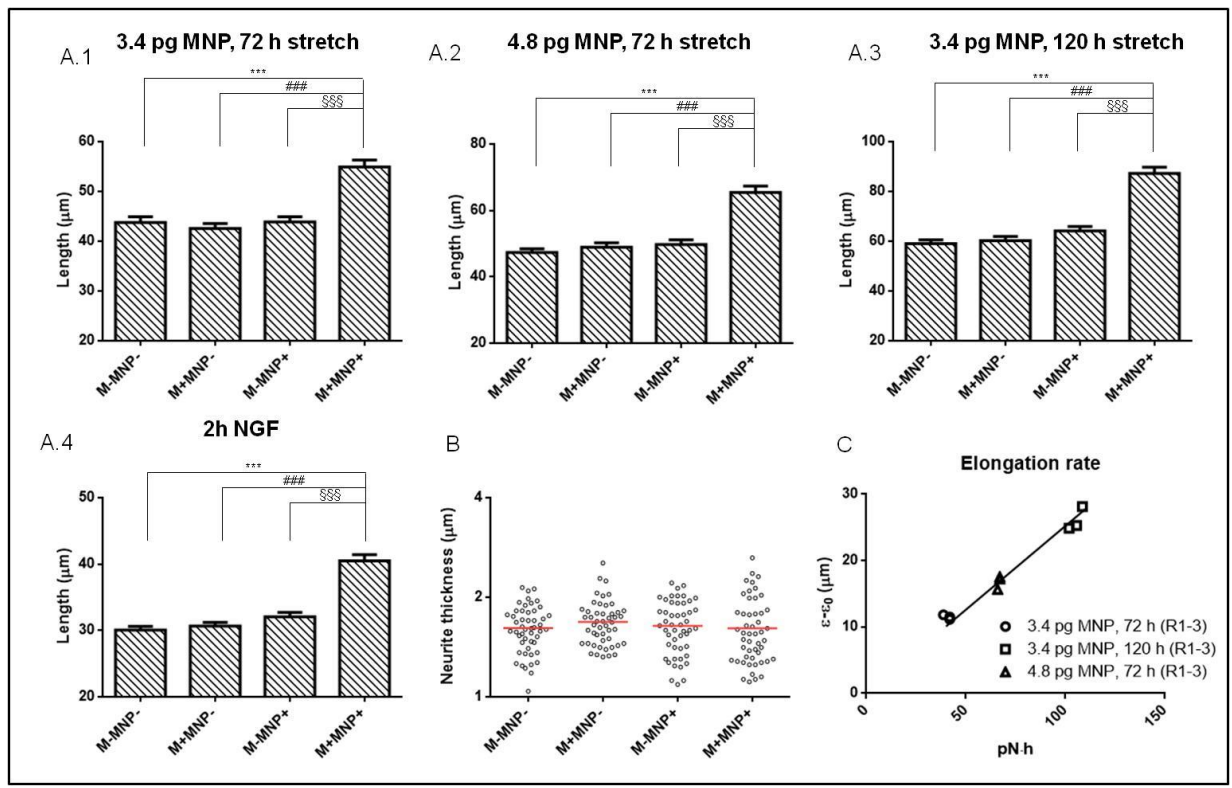

**Fig. 2**. pN-stretching of PC12 cell neurites. A1-4) Neurite length for stretching condition '3.4 pg MNP, 72h', '3.4pg MNP, 120h', '4.8pg MNP, 72h' and "2h NGF", respectively. n=600. Kruskal Wallis test, followed by HDS correction: $p=9.3 \cdot 10^{-20}$ (A1), $p=1.9 \cdot 10^{-20}$ (A2), $p=8.0 \cdot 10^{-25}$ (A3) and $p=7.5 \cdot 10^{-31}$ (A4). B) Neurite thickness (condition '4.8pg MNP, 72h'), n=50. Kruskal Wallis test. p=0.43. C) Differential elongation versus applied force per time. The applied force was calculated according to the mathematical model (supplementary information, S2). The differential elongation is expressed as the difference of elongation between the stretched and non-stretched conditions. The elongation rate ($0.252 \pm 0.0053 \mu m h^{-1} pN^{-1}$) was calculated by linear regression analysis (95% of confidence level, p<0.0001).

We also evaluated if the pN-stretching could initiate neurite formation. Data analysis showed no difference in the number of neurites per cell among the groups. However, there is a trend for an increase in the stretched samples, sometimes reaching a weak statistical significance when compared to some control groups (supplementary information, S4). We also found that the pN-stretching is not per se a signal sufficient to sustain PC12 differentiation (and neurite initiation) in absence of NGF. However, by performing a short-term exposure of the cultures to NGF (2 h incubation), stretch-growth was observed in all 3 replicates (table1, 'NGF 2h' and Fig. 2.A4) and, similarly to the conditions of cells continuously exposed to NGF, the length increase was significant only for stretched neurites.

We performed RNAseq of MNP-labelled cells in stretched versus non-stretched conditions and we did not found gene expression dysregulation (Fig. 3.A), confirming that the two samples were identical (Fig. 3.B) and excluding cytotoxicity or involvement of nuclear mechanotransdution. Indeed, local mechanisms triggered by whole neurite stretching would be responsible for neurite elongation by mass addition.

The data obtained by these experiments strongly support the conclusion that stretch-growth of neurites is not threshold-dependent, in line with recent models, which do not assume or require a force threshold involved in stretch growth of neurite (Recho et al., 2016).

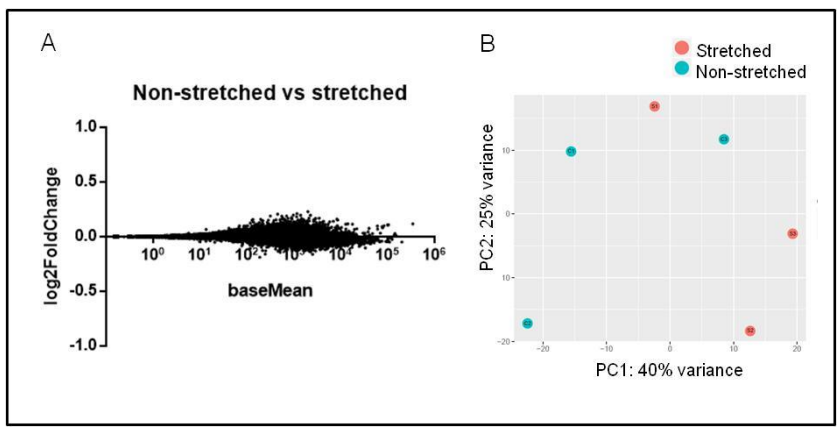

**Fig. 3.** A) Logarithmic fold change of gene expression (stretched versus non-stretched condition). A value of ±1 (corresponding to a 2-fold increase) is typically considered a reasonable cut off for gene dysregulation. A) PCA plot. There is no separation between groups, which confirms that stretched and non-stretched samples are identical. The stretched condition is $M^+MNP^+$, the non-stretched condition is $M^-MNP^+$.

**Materials and methods**

*Magnetic nanoparticles*

MNPs used in this work are polyethyleneimine (PEI, 25 kDa) $Fe_3O_4$ nanoparticles, which we have extensively characterized elsewhere (Calatayud et al., 2013). According to our previous characterization (Calatayud et al., 2013), we used the following data for the mathematical model: size 25nm, saturation magnetization 58Am$^2$kg and density 5·10$^3$kgm$^3$.

*Cell cultures*

Rat pheochromocytoma PC12 cells obtained from American Type Culture Collection (ATCC) were cultured in Dulbecco's modified Eagle's medium with 10% horse serum, 5% fetal bovine serum (FBS), 100IUml$^{-1}$ penicillin, 100µgml$^{-1}$ streptomycin and 2mM L-glutamine. Cells were cultured in Petri dishes coated with poly-l-lysine (PLL, Sigma, P1274) and maintained at 37°C in a saturated humidity atmosphere of 95% air and 5% $CO_2$. For cell differentiation, PC12 cells were incubated in serum-reduced media (1% FBS). Experiments were performed at low density, i.e. 2.5·10$^4$ cells per cm$^2$. Cells were used at passages 6-12. Microscopy and digital image acquisitions were carried out with an Olympus 1X71/1X51 inverted microscope.

*Cell uptake*

The amount of MNPs in cells was quantified by using the thiocyanate assay, according to a protocol we already published (Calatayud et al., 2013). Briefly, cell pellet was re-suspended in 50µl of a solution of 6M HCl: 65% $HNO_3$ v/v and incubated at 60°C for 1h. The sample was water diluted 1:10, an equal volume of 1.5M KSCN was added and absorbance recorded at 478nm. The calibration curve was y = 0.0172x + 0.0015 ($R^2$ = 1) where y is the absorbance @478 nm and x is the amount of MNPs (µg).

*Cell viability*

Cells were incubated for 72h with MNPs. Then, cells were incubated with 1µM Hoechst for 10 minutes at 37°C and with 10µgml$^{-1}$ propidium iodide (PI) for additional 5 minutes at 37°C. For each sample, the number of necrotic and pyknotic cells was counted on a random population of 1000 cells. For the evaluation of the cell doubling time, cells were removed by trypsinization after 48h ($t_0$) or 96h ($t_1$) of incubation with the particles and counted in a Burker's chamber. Cell doubling time ($T_d$) was calculated by using the following formula: $T_d = (t_1-t_0)\cdot \ln(2)/\ln(q_1/q_0)$, with $q_0$ and $q_1$ the cell number at time $t_0$ and $t_1$, respectively.

*Magnetic field*

Experiments were carried out in 35mm Petri dishes placed inside a halbach-like cylinder magnetic applicator, which provided a constant magnetic field gradient of 46.5Tm$^{-1}$ in the radial direction outwards (Riggio et al., 2014).

*Stretching assay*

PC12 cells were seeded in 35mm dishes pre-coated with $1\mu gml^{-1}$ PLL. 24 hours later the induction of differentiation (i.e. incubation in reduced serum medium supplemented with 100ngml$^{-1}$ NGF), the Petri dish was put inside the magnetic applicator. Analysis was performed by using "image analysis software "Image J" (http://rsb.info.nih.gov/ij/). Neurite length *l* was evaluated by using the plugin "Neuron J" and 200 neurites were analysed from 10x magnification images (randomly acquired). For the analysis, a cut-off of 10µm in length was fixed and neurites in networks were excluded. The longest path was measured for branched neurites. For neurite thickness, a population of 100 neurites was analyzed from 20x magnification images (randomly acquired). For each neurite, the thickness *s* was calculated as *s=A/l*, being *A* the neurite area that was precisely calculated from images after threshold normalization, binary conversion and elimination of elements with size below the cut-off. Cell sprouting was calculated by counting the number of processes coming out from isolated cells (n=100).

The volume of cell cytoplasm was calculated by acquiring 60X images in a population of suspended (Hoechst stained) cells, measuring cellular and nuclear diameter (n=25).

*RNAseq*

For RNAseq experiment, 9 h later the application of the magnetic field, samples were chilled in $N_2$ and the RNA was extracted with RNeasy Kit (Qiagen), according to the manufacturer's instructions. Quality check (QC) was performed (RIN=10). RNAseq was performed with the platform Illumina NextSeq500. RNA library was prepared by using polyA selection library and sequencing mode: PE 2x75bp, 25-40M. RNAseq and data analysis were performed at Glasgow Polyomics, UK.

*Electron microscopy*

SEM/FIB cross sectioned cells were performed using scanning electron microscopy (SEM INSPECT F50, FEI Company) and dual-beam FIB/SEM (Nova 200 NanoLab, FEI Company). PC12 cells were grown on coverslip coated with PLL and treated with MNPs (10 μgml$^{-1}$). After 24 h of incubation the cells were washed with PBS, fixed and dehydrate. After drying the samples were sputtered with 30 nm of gold. SEM images were taken at 5 and 30 kV with a FEG column, and a combined Ga-based 30 kV (10pA) ion beam was used to cross-section single cells. These investigations were completed by EDX for chemical analysis.

*Statistical analysis*

Data were plotted with GraphPad Software, version 6.0. Values are reported as the mean ± standard error of the mean. Data distributions were analyzed by Kolmogorov-Smirnov test. Statistical significance was assessed by one-way analysis of variance. Specifically, for non-normal data distribution, we used Kruskal-Wallis analysis, followed by multi-compare analysis (95% confidence) whereas for normal data distributions, we used Anova followed by Bonferroni correction. Significance was set at $p \leq 0.05$. Statistical analysis were performed in Matlab R14 workspace (functions "test2","kstest", "anova1", "bonferroni", "multicompare") or with GraphPad Software, version 6.0.

**Supplementary Information**

*S1. Cell viability*

We tested particle toxicity in order to exclude batch-dependent toxicity. Experimental data confirm previous observations. PC12 cells were incubated with 0, 10, 20, 40 or 60µgml$^{-1}$ MNPs for 72h. Cell viability was tested by using PI dye exclusion assay. We found the percentage of PI positive cells below 5% for all groups, with a statistically significant variation from the control only for the highest concentration of 60µgml$^{-1}$ (p=0.01) (Fig. S1.1). The percentage of apoptotic nuclei was found below 4% for all groups without any significant difference among groups (p=0.23) (Fig. S1.1). Although nanomaterials do not alter cell viability, they can delay cell cycle progression when mild damages are induced. The evaluation of the cell doubling time is thus a sensitive parameter to predict nanotoxicity. However, we did not detect any change in cell doubling time, even at the highest dose, with no difference among groups (p=0.95) (Fig. S1.2). We concluded that particles can be safely administered to cells and experiments were performed by using the lowest concentration tested, i.e. 10µgml$^{-1}$.

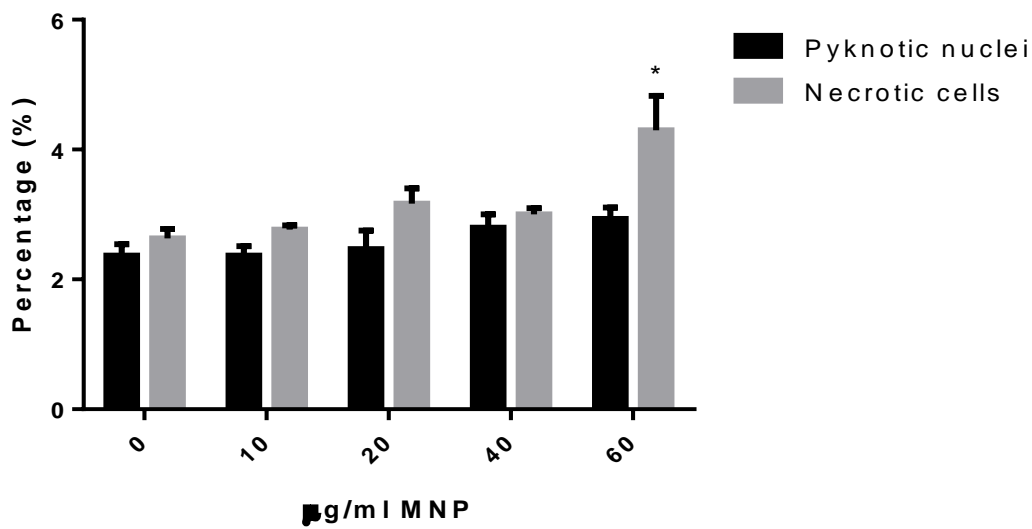

**Fig. S1.1.** Cell viability was assessed on PC12 cells after 72h of incubation with MNPs. Necrosis was calculated as percentage of PI positive cells. 1-way ANOVA test followed by Bonferroni correction, n=3, p=0.01. Pyknotic nuclei were evaluated by DAPI staining. ANOVA test followed by Bonferroni correction, n=3, p=0.23.

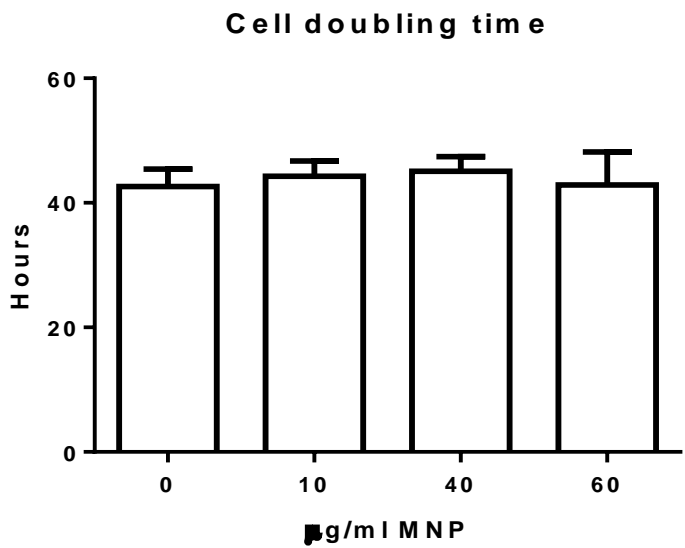

**Fig. S1.2.** Cell doubling time. n=3, 1-way ANOVA test, p=0.95.

*S2. Mathematical model*

Our experimental set-up has been designed to apply a constant force in any point of the Petri dish, with a radial direction outwards. The only non-null component of the magnetic field gradient is thus the radial one (dB/dr=46.5 Tm$^{-1}$) (Fig. S2.A). As our particles have a saturation magnetization M$_s$ of 58 Am$^2$kg$^{-1}$ and a coercive field H$_c$ of 4.81 kAm$^{-1}$ (Calatayud et al., 2013), we can assume that particle magnetization saturated and the magnetic force acting on the single particle is given by:

$$F = m_s \frac{dB}{dr} = \rho V M_s \frac{dB}{dr} \qquad (1)$$

where $\rho$ is the particle density and V the particle volume. The resulting force F acting on the single particle is $1.1 \cdot 10^{-16}$N. The mechanical force acting on the neurite thus depends on the number of particles inside the neurite. We made the simplest assumption that particles have a uniform distribution in cell cytoplasm, included the neurite. This assumption is strongly corroborated by previous analysis. Cells were sectioned from the cell body to the growth cone and microanalysis revealed the presence of Fe in any section, excluded the fractions occupied by nucleus (Riggio et al., 2014). A single neurite will be thus subjected to a force F$_{neur}$ given by:

$$F_{neur} = n \left(\frac{V_{neur}}{V_{cyt}}\right) F \qquad (2)$$

where $n$, V$_{neur}$, V$_{cyt}$ are the number of particles in the cell, the neurite volume and the cytoplasm volume, respectively.

In a polar coordinate system, where the origin is the neurite, the radial coordinate $r$ is the neurite direction and the angular coordinate is $\theta$, this force has two components, the radial $F_{neur,r} = F_{neur}\cos\theta$ and the angular $F_{neur,\theta} = F_{neur}\sin\theta$, being $\theta$ the angle between the direction of the magnetic force and the direction of the neurite (Fig. S2.B). The first component is responsible for stretching the neurite along its length (the latter is the component orienting the neurite). The present work has been performed with a statistical approach querying a neurite population (n=200). In this context, the mean force exerted on the neurite is given by:

$$|\overline{F_{neur}\cos\theta}| = F_{neur}|\overline{\cos\theta}| \qquad (3)$$

All parameters in equations 1-3 have been evaluated experimentally or extrapolated by experimental data. Specifically $n$ has been extrapolated by data fitting provided in Fig. S3 (each particle has an average volume and weight of $8.1 \cdot 10^{-24}$m$^3$ and $4.1 \cdot 10^{-5}$pg, respectively). V$_{cyt}$ was calculated to be 542±102µm$^3$ (n=25). $V_{neu}$ was calculated by modeling the neurite as a cylinder, being length and thickness distributions known from Fig. 1. Finally, the angle distribution in this configuration has been deeply characterized and modeled in our previous work (Riggio et al., 2014). Specifically, we found that angles are randomly distributed in non-stretched conditions (i.e. $\overline{|\cos\theta|}=1/2$) but we found that neurites preferentially align along the direction imposed by the magnetic field in stretched condition (i.e. $\overline{|\cos\theta|}>1/2$) and we used the mathematic model already described (Riggio et al., 2014) to calculate the corresponding value.

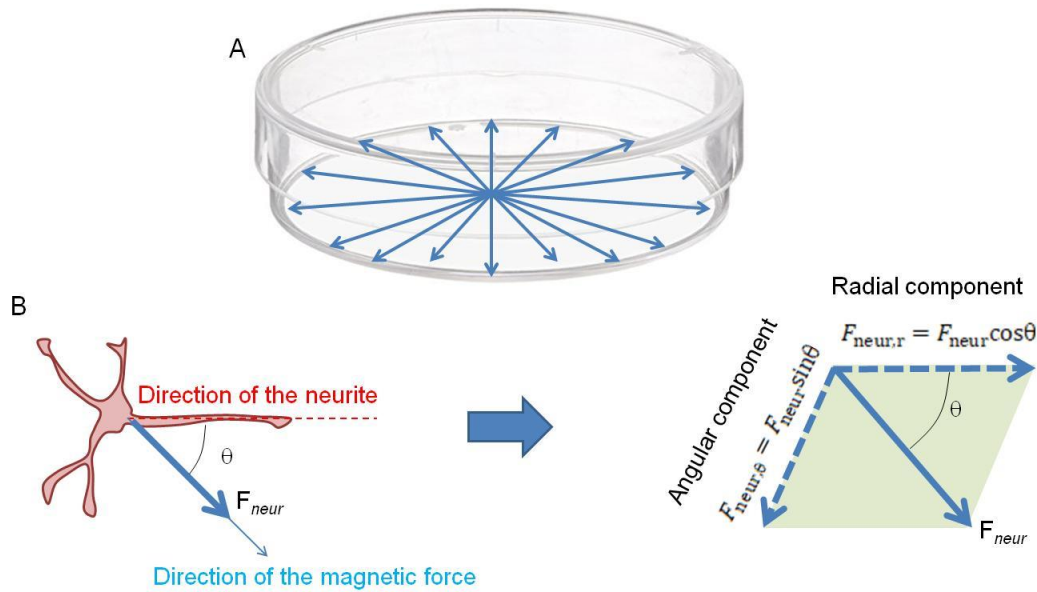

**Fig. S2.** Illustration of the force acting on the neurite. A) The magnetic field gradient and the magnetic force acting on the single MNP are constant in amplitude and the direction is radial outwards. B) The MNPs entrapped in the neurite exert a magnetic force. The radial component of this force is responsible for stretching the neurite. The angular component is responsible for changing the orientation of the neurite. Depending on the orientation of the neurite with respect to the orientation of the magnetic field, the radial component could be directed from soma to tip (in the example) or from tip to soma.

*S3. Cell labelling with MNPs*

We performed 2 different procedures of cell labelling with MNPs. In the first procedure, MNPs were added to the differentiation medium. 96h after differentiation, the amount of particles was calculated as described in the M&M section and normalized per the cell number. The amount of particles interacting with cells in this experimental configuration was estimated to be 3.45pg of MNPs (corresponding to $8.44 \cdot 10^4$ particles) per cell. The value was extrapolated by a dose-response assay, which showed an excellent data correlation (n=6, $R^2$=0.94) (Fig. S3).

In the second procedure, MNPs were added to the cell growth medium and 48h later the medium was removed and replaced with the differentiation medium. Similarly, 96h after differentiation, the amount of particles was calculated as described in the M&M section and normalized per the cell number. The amount of particles interacting with cells in this new experimental configuration was 4.85 pg of MNPs (corresponding to $1.19 \cdot 10^5$ particles) per cell. Similarly, the value was extrapolated by a dose-response assay (n=6, $R^2$=0.99) (Fig. S3).

As expected, cells cultured in the cell growth medium showed a greater ability to internalize MNPs than cells cultured in the differentiation medium.

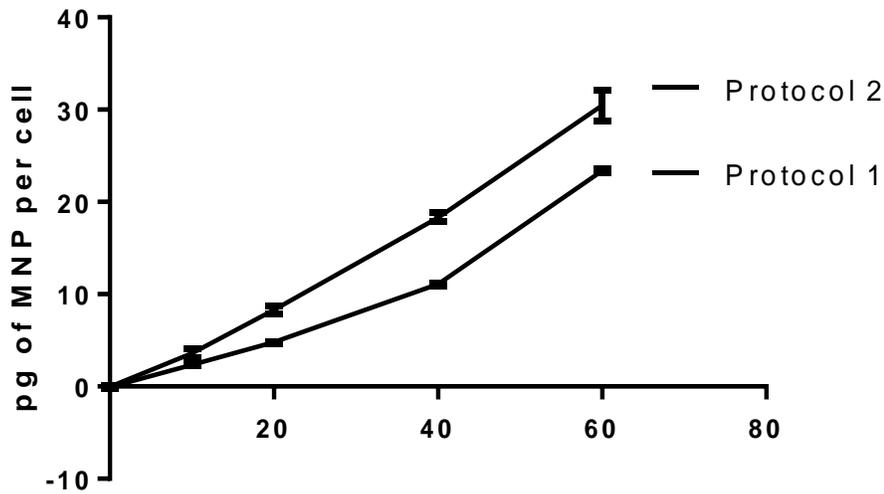

**Fig. S3.** MNP cell labelling. Protocol 1: cells cultured for 96h in differentiation medium supplemented with MNPs. Protocol 2: cells cultured for 48h in cell growth medium supplemented with MNPs, followed by 96 hours in differentiation medium.

*S4. Cell sprouting*

**Table S**. Neurite number per cell.

|  | M$^-$MNP$^-$ | M$^+$MNP$^-$ | M$^-$MNP$^+$ | M$^+$MNP$^+$ | P value |
|---|---|---|---|---|---|
| 3.4pg MNP, 72h stretch | 2.34±0.12 | 2.27±0.11 | 2.270±0.11 | 2.47±0.13 | 0.765 |
| 4.8pg MNP, 72h stretch | 2.36±0.10 | 2.42±0.12 | 2.30±0.11 | 2.58±0.12 | 0.490 |
| 3.4pg MNP, 120h stretch | 3.12±0.12 | 2.97±0.12 | 3.22±0.13 | 3.65±0.13$^{*,\#\#}$ | 0.001 |
| NGF 2h | 2.02±0.10 | 2.34±0.12 | 2.14±0.11 | 2.45±0.11$^*$ | 0.027 |

N=100. Kruskal Wallis test followed by HDS correction.

## Acknowledgments

We thank Prof J. Lambert for his assistance and support he gave for discussions. The study was partially supported by EU FP7, IEF Marie Curie (MECAR, 622122) and Wings for Life Foundation.

## Author contributions

V.R. and A.C. conceived and designed the study. V.R. and F.F. performed the experiments and analyzed data. M.P.C and G.F.G performed MNP synthesis, characterization and electron microscopy analysis. V.R and A.C. wrote the manuscript.

## References

Athamneh, A.I., and D.M. Suter. 2015. Quantifying mechanical force in axonal growth and guidance. *Frontiers in cellular neuroscience*. 9:359.
Breau, M.A., I. Bonnet, J. Stoufflet, J. Xie, S. De Castro, and S. Schneider-Maunoury. 2017. Extrinsic mechanical forces mediate retrograde axon extension in a developing neuronal circuit. *Nature communications*. 8:282.
Calatayud, M.P., C. Riggio, V. Raffa, B. Sanz, T.E. Torres, M.R. Ibarra, C. Hoskins, A. Cuschieri, L. Wang, J. Pinkernelle, G. Keilhofff, and G.F. Goya. 2013. Neuronal cells loaded with PEI-coated Fe3O4 nanoparticles for magnetically guided nerve regeneration. *J Mater Chem B*. 1:3607-3616.
Calatayud, M.P., B. Sanz, V. Raffa, C. Riggio, M.R. Ibarra, and G.F. Goya. 2014. The effect of surface charge of functionalized Fe3O4 nanoparticles on protein adsorption and cell uptake. *Biomaterials*. 35:6389-6399.
Chada, S., P. Lamoureux, R.E. Buxbaum, and S.R. Heidemann. 1997. Cytomechanics of neurite outgrowth from chick brain neurons. *Journal of Cell Science*. 110:1179-1186.
Dennerll, T.J., H.C. Joshi, V.L. Steel, R.E. Buxbaum, and S.R. Heidemann. 1988. Tension and compression in the cytoskeleton of PC-12 neurites. II: Quantitative measurements. *The Journal of cell biology*. 107:665-674.
Dennerll, T.J., P. Lamoureux, R.E. Buxbaum, and S.R. Heidemann. 1989. The Cytomechanics of Axonal Elongation and Retraction. *Journal of Cell Biology*. 109:3073-3083.
Fass, J.N., and D.J. Odde. 2003. Tensile force-dependent neurite elicitation via anti-beta 1 integrin antibody-coated magnetic beads. *Biophysical Journal*. 85:623-636.
Franze, K., and J. Guck. 2010. The biophysics of neuronal growth. *Reports on Progress in Physics*. 73.
Lamoureux, P., Z.F. AltunGultekin, C.J. Lin, J.A. Wagner, and S.R. Heidemann. 1997. Rac is required for growth cone function but not neurite assembly. *Journal of Cell Science*. 110:635-641.
Lamoureux, P., R.E. Buxbaum, and S.R. Heidemann. 1989. Direct evidence that growth cones pull. *Nature*. 340:159-162.
Lamoureux, P., S.R. Heidemann, N.R. Martzke, and K.E. Miller. 2010. Growth and elongation within and along the axon. *Developmental neurobiology*. 70:135-149.
Miller, K.E., and M.P. Sheetz. 2006. Direct evidence for coherent low velocity axonal transport of mitochondria. *The Journal of cell biology*. 173:373-381.
O'Toole, M., P. Lamoureux, and K.E. Miller. 2008a. A physical model of axonal elongation: Force, viscosity, and adhesions govern the mode of outgrowth. *Biophysical Journal*. 94:2610-2620.
O'Toole, M., R. Latham, R.M. Baqri, and K.E. Miller. 2008b. Modeling mitochondrial dynamics during in vivo axonal elongation. *Journal of theoretical biology*. 255:369-377.

Pinkernelle, J., V. Raffa, M.P. Calatayud, G.F. Goya, C. Riggio, and G. Keilhoff. 2015. Growth factor choice is critical for successful functionalization of nanoparticles. *Front Neurosci-Switz*. 9.

Recho, P., A. Jerusalem, and A. Goriely. 2016. Growth, collapse, and stalling in a mechanical model for neurite motility. *Physical review. E*. 93:032410.

Riggio, C., M.P. Calatayud, M. Giannaccini, B. Sanz, T.E. Torres, R. Fernandez-Pacheco, A. Ripoli, M.R. Ibarra, L. Dente, A. Cuschieri, G.F. Goya, and V. Raffa. 2014. The orientation of the neuronal growth process can be directed via magnetic nanoparticles under an applied magnetic field. *Nanomed-Nanotechnol*. 10:1549-1558.

Riggio, C., S. Nocentini, M.P. Catalayud, G.F. Goya, A. Cuschieri, V. Raffa, and J.A. del Rio. 2013. Generation of Magnetized Olfactory Ensheathing Cells for Regenerative Studies in the Central and Peripheral Nervous Tissue. *Int J Mol Sci*. 14:10852-10868.

Ruthel, G., and P.J. Hollenbeck. 2000. Growth cones are not required for initial establishment of polarity or differential axon branch growth in cultured hippocampal neurons. *The Journal of neuroscience : the official journal of the Society for Neuroscience*. 20:2266-2274.

Suter, D.M., and K.E. Miller. 2011. The emerging role of forces in axonal elongation. *Progress in neurobiology*. 94:91-101.

Zheng, J., P. Lamoureux, V. Santiago, T. Dennerll, R.E. Buxbaum, and S.R. Heidemann. 1991. Tensile Regulation of Axonal Elongation and Initiation. *Journal of Neuroscience*. 11:1117-1125.